\documentclass[10pt,twocolumn]{article}
\renewcommand{\thesection}{\Roman{section}}
\usepackage{graphicx}
\graphicspath{{figures/}}
\usepackage{amsmath}
\usepackage{amssymb}
\usepackage[a4paper,margin=.7in]{geometry}
\usepackage{amsmath}
\usepackage{graphicx}
\usepackage{booktabs}
\usepackage{enumitem}
\usepackage[hidelinks]{hyperref}
\usepackage{graphicx}
\usepackage{booktabs}
\usepackage{enumitem}
\usepackage{hyperref}
\usepackage{cite}
\usepackage{xcolor}
\usepackage{titlesec}
\usepackage{array}
\usepackage{wrapfig}
\usepackage{placeins}

\titleformat{\section}
{\normalfont\large\bfseries}
{\thesection.}{1.em}{}

\title{Indirect-Drive Fusion Target Design for Commercial Fusion Energy}
\author{
\parbox{0.95\textwidth}{\centering
C. R. Weber$^{*}$, S. Bhandarkar, T. Briggs, T. Chapman, T. M. Fears, B. A. Hammel,\\ D. D.-M Ho,
O. Hurricane, B. Kozioziemski, J. Milovich, A. Nikroo, A. Oudin,\\ W. Riedel, A. Wray\\
\textit{Lawrence Livermore National Laboratory}\\[0.8em]
A. L. Kritcher$^{\dagger}$, N. Alexander, G. Cearley, M. Dunne, J. Gaffney,\\
D. Hammond, J. Kilkenny, R. Lau, J. Lawson, J. Ludwig, B. MacGowan,\\
R. Peterson, V. Smalyuk, R. Toro, and the Inertia Collaboration$^{\ddagger}$\\
\textit{Inertia Enterprises}\\ [0.8em]
{\small
Corresponding authors: $^{*}$weber30@llnl.gov and $^{\dagger}$annie@inertia.com\\
$^{\ddagger}$See Acknowledgements for the Inertia Collaboration author list 
}
}
}
\date{\today}

\begin{document}
\twocolumn[{
\begin{@twocolumnfalse}

\maketitle


\begin{abstract}
This paper presents the physics basis for commercially relevant laser indirect-drive (LID) (radiation-driven) inertial fusion energy (IFE) using a 10 MJ laser driver. To date, this approach, proven at the National Ignition Facility (NIF), remains the first and only controlled fusion method to demonstrate the key physics required for fusion energy production, including a self-sustained burning plasma, substantially de-risking the path to commercial fusion energy.  Building directly on these results, we present scaled designs to larger target sizes and fusion gains relevant for commercial power generation ($G\sim 26$--$43$).  The designs remain close to experimentally demonstrated ignition physics, modifying target components to improve scalability, manufacturability, and cost-effectiveness for fusion energy applications while preserving ignition-relevant implosion physics and fusion power plant compatibility.  The baseline platform retains a high-density carbon ablator and clean cryogenic DT fuel layering while extending ignition platforms to substantially larger fuel masses (exceeding 10 times that of current ignition experiments at the NIF) and higher burn fractions ($\sim$40\%) with total areal densities at stagnation of $\sim$3~g/cm$^2$. Benchmarked simulations anchored to NIF ignition experiments (using HYDRA and LASNEX) predict that these designs achieve robust ignition and propagating burn at substantially higher fusion yields (265--427~MJ) with significant ignition margin (2--4$\times$ relative to NIF) against hydrodynamic instabilities and representative power-plant non-idealities, including low-mode asymmetry, polycrystalline DT ice roughness, HDC ablator voids, and target-support and fill-hole perturbations. We also show that implosion symmetry and laser-plasma interactions (LPI) can be controlled with our novel multi-beam configuration using thousands of laser beam-lines. 
\end{abstract}
\vspace{0.5cm}
\end{@twocolumnfalse}

}]

\section{Introduction}
Over the past several years, key physics milestones required for fusion energy production have been experimentally demonstrated, including a burning self-heated plasma \cite{Kritcher_NP_2021,Zylstra_Nature_2021}, Lawson's criterion for ignition \cite{Kritcher_0808_2022,Ign_0808_2022,Zylstra_0808_2022}, and target gain exceeding unity \cite{kritcher2023,kritcher2023pop,AbuShawareb2024PRL,Pak2024PRE,hurricane_2023}. These milestones were achieved using the Laser Indirect-Drive (radiation-driven) inertial confinement fusion (LID-ICF) approach \cite{Nuckolls_Nature_1972,Kidder_NF_1974,Clarke_PRL_1973,Lindl_PoP_1995,Lindl_PoP_2004,Haan_PoP_2011} with 1.9--2.05~MJ of laser energy \cite{JM_1,Di_Nicola__NF_2018} and the ``Hybrid-E'' design platform \cite{Kritcher_PoP_2020,Kritcher_PoP_2021,Kritcher_NP_2021,Kritcher_0808_2022,kritcher2023,Zylstra_PoP_2020,Zylstra_PRL_2021,Zylstra_Nature_2021,Ign_0808_2022,Zylstra_0808_2022,main_2023,zylstra_2023}. To date, LID-ICF is the only approach to have demonstrated these physics milestones. Reaching this level of performance required decades of development, substantial government research investment, and the resolution of numerous unforeseen technical challenges \cite{Hurricane_RMP_2023,Rosen_PoP_2024,Edwards_RMPP_2025,Lindl_Haan_PoP_2026}. Collectively, these results provide a mature and experimentally validated foundation on which to build toward fusion energy production.

Experiments performed at the National Ignition Facility (NIF) \cite{Moses_JoP_2016} also achieved a Lawson parameter \cite{Lawson_1957} more than an order of magnitude higher than any other fusion concept demonstrated to date \cite{Wurzel}, placing laser indirect-drive with solid-state lasers in a significant lead toward practical fusion energy.  The Lawson parameter ($nT\tau$) is a key figure of merit in fusion research because it characterizes the combination of plasma density ($n$), temperature ($T$), and confinement time ($\tau$) required for fusion energy generation to exceed energy losses. As such, it provides a quantitative metric for comparing the proximity and viability of different fusion concepts for net energy production, i.e., more energy generated from fusion than required to heat and confine the fusion plasma.

The significance of these results extends beyond the achievement of ignition itself. The NIF experiments demonstrated that radiation-driven implosions with solid-state laser drivers \cite{JM_1,Di_Nicola__NF_2018} can access and control the narrow region of parameter space required for ignition, including symmetry \cite{Kritcher_PoP_2014,Kritcher_PRE_2018,Callahan_PoP_2018,MacGowan_HEDP_2020,Casey_PRL_2021}, adiabat, implosion velocity, instability growth, and laser-plasma interactions. The indirect-drive approach naturally smooths driver asymmetries through x-ray conversion within the hohlraum, reducing mid and high-mode imprint on the capsule. 
In contrast, laser direct-drive could offer the possibility of improved coupling efficiency, but will first need to demonstrate mitigation of laser imprint, cross-beam energy transfer, and other laser-plasma-interaction (LPI) effects that can seed hydrodynamic perturbations \cite{Schmitt_PoP_2001}, reduce the coupled drive \cite{Radha_JPCS_2016_DirectDrive}, and preheat the fuel \cite{Smalyuk_PRL_2008}. These effects leave significant uncertainties in extrapolating direct-drive to ignition and burn-propagation regimes \cite{Lees_PoP_2025}.

Beyond its demonstrated physics advantages, indirect drive is also well suited for fusion power plant integration. The two-sided laser geometry leaves sufficient solid angle for tritium breeding in a lithium blanket through neutron-induced reactions (${}^{7}\mathrm{Li} + n \rightarrow {}^{4}\mathrm{He} + T + n - 2.5~\mathrm{MeV}$ and ${}^{6}\mathrm{Li} + n \rightarrow {}^{4}\mathrm{He} + T + 4.8~\mathrm{MeV}$), while enabling polar neutron shielding of the target injection system; the hohlraum itself also helps shield the first wall from the prompt x-ray pulse and charged-particle debris. In addition, target designs and their components can be iterated and replaced without rebuilding the laser facility, as repeatedly demonstrated at NIF. In contrast, magnetic confinement concepts are largely constrained by reactor geometry and magnetic configuration, making major performance advances dependent on costly modifications to existing facilities or the construction of entirely new, and typically larger, systems.

With the physics basis for ignition experimentally established, the central challenge for commercial fusion energy is achieving robust and repeatable burn propagation in the presence of non-idealities expected in a fusion power plant environment. These include imperfections associated with low-cost target materials, manufacturing defects, imperfect cryogenic fuel layering, target modifications for survival, and off-center target positioning during injection and tracking. In addition, although ignition has now been reproducibly demonstrated on the NIF with fusion yields $\sim$8~MJ and gain $\sim$4$\times$, practical fusion energy production with sufficient net wall-plug gain requires higher yields approaching hundreds of MJ, higher repetition rates, cheaper targets, and laser drivers significantly more efficient than NIF, which operates at $<1\%$ wall-plug efficiency.

In this work, we show that this regime is accessible with an LID-ICF platform through an increase in laser driver energy by $\sim$5$\times$ (10~MJ), an increase in fuel mass and coupled energy by 9--15$\times$ and 10--13$\times$, respectively, and the use of high-repetition-rate ($\sim$10 Hz) diode-pumped solid-state laser systems capable of improving wall-plug efficiency to $\sim$10--20\%, more than an order of magnitude above the NIF laser architecture.
The baseline platform is intentionally designed around well-understood physics and benchmarked design strategies, moderate increases in target gain compared to direct scaling from current experiments, and experimentally demonstrated laser technologies rather than relying on higher-risk concepts in new target designs or laser technologies. The strategy is to establish a robust 10 MJ-class First-of-a-Kind Fusion Pilot Plant design with substantial driver margin, providing a foundation for subsequent optimization of target and laser designs to reduce cost and improve performance for n$^{\mathrm{th}}$-of-a-kind systems.

We also show that a key advantage of moving to substantially larger driver energies and fuel masses is that the implosion operates farther from the ignition threshold, providing approximately 2--4$\times$ greater modeled ignition margin than NIF, reducing sensitivity to target imperfections, laser delivery variations, and other power-plant-relevant perturbations while maintaining high fusion energy production (265--427~MJ).  These designs achieve substantially higher total areal densities (2.6--3.3~g/cm$^2$) and burn fractions (37--41\%) through increased fuel and capsule mass rather than higher implosion convergence, on which most high-gain concepts rely despite increased sensitivity to perturbations and less well-understood performance \cite{NIC_Report_2011,Atzeni_PoP_1999,Azechi_NF_2013}.

This paper is organized as follows. Section \ref{Section2} describes the LID approach and scaling to high drive energy platforms. Section \ref{Section3} presents the baseline target design, including the target configuration, laser drive configuration, and integrated performance characteristics. Section  \ref{Section4} examines target robustness and sensitivity to realistic perturbations relevant to fusion energy applications, including material changes, target and fuel defects, asymmetry, and injection-related non-idealities. Section \ref{Section5} presents reactor-relevant gain and power production, and Section \ref{Section6} presents the conclusions. Additional details regarding the integrated design methodology and modeling framework are provided in the Appendix.

 \begin{figure}[t]
\centering
\includegraphics[width=1.0\linewidth]{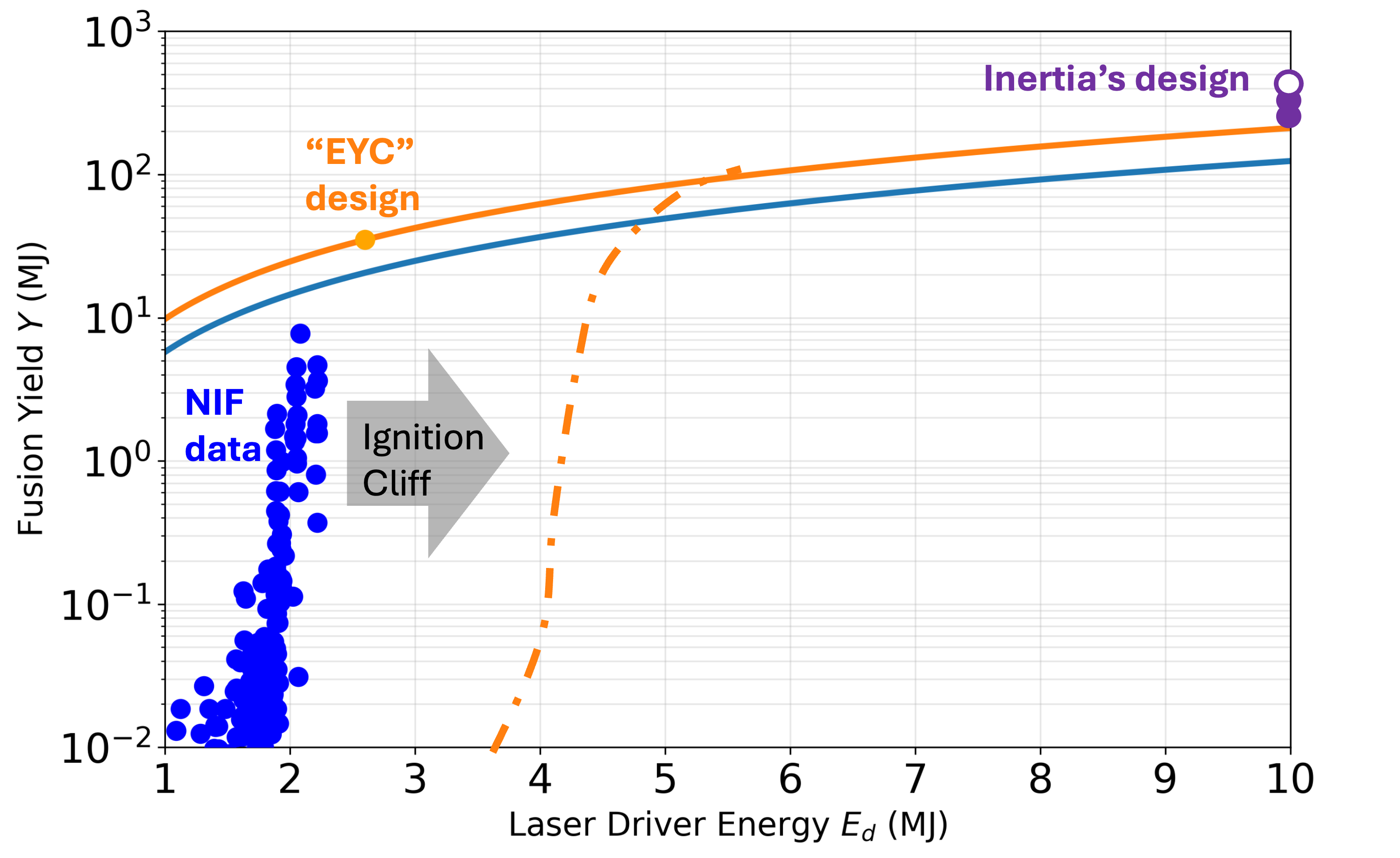}
\caption{Fusion energy produced (yield) as a function of laser driver energy ($E_d$). Overlaid are experimental data points from NIF (blue), Meyer--ter--Vehn scaling extrapolations (orange curve) from the EYC 2.6 MJ design point (orange point) \cite{10.1117/12.3056940}, and Meyer--ter--Vehn scaling from current NIF ignition experiments assuming idealized 1D yield of 15~MJ at 2.08 MJ $E_d$ (blue curve).
Inertia’s 10 MJ-class target designs span 265--427~MJ. The dashed orange curve illustrates how the effective ignition cliff may shift to higher required driver energies as realistic fusion power plant non-idealities are introduced.}
\label{Fig1}
\end{figure}

\begin{figure*}[t]
\centering
\includegraphics[width=1.\linewidth]{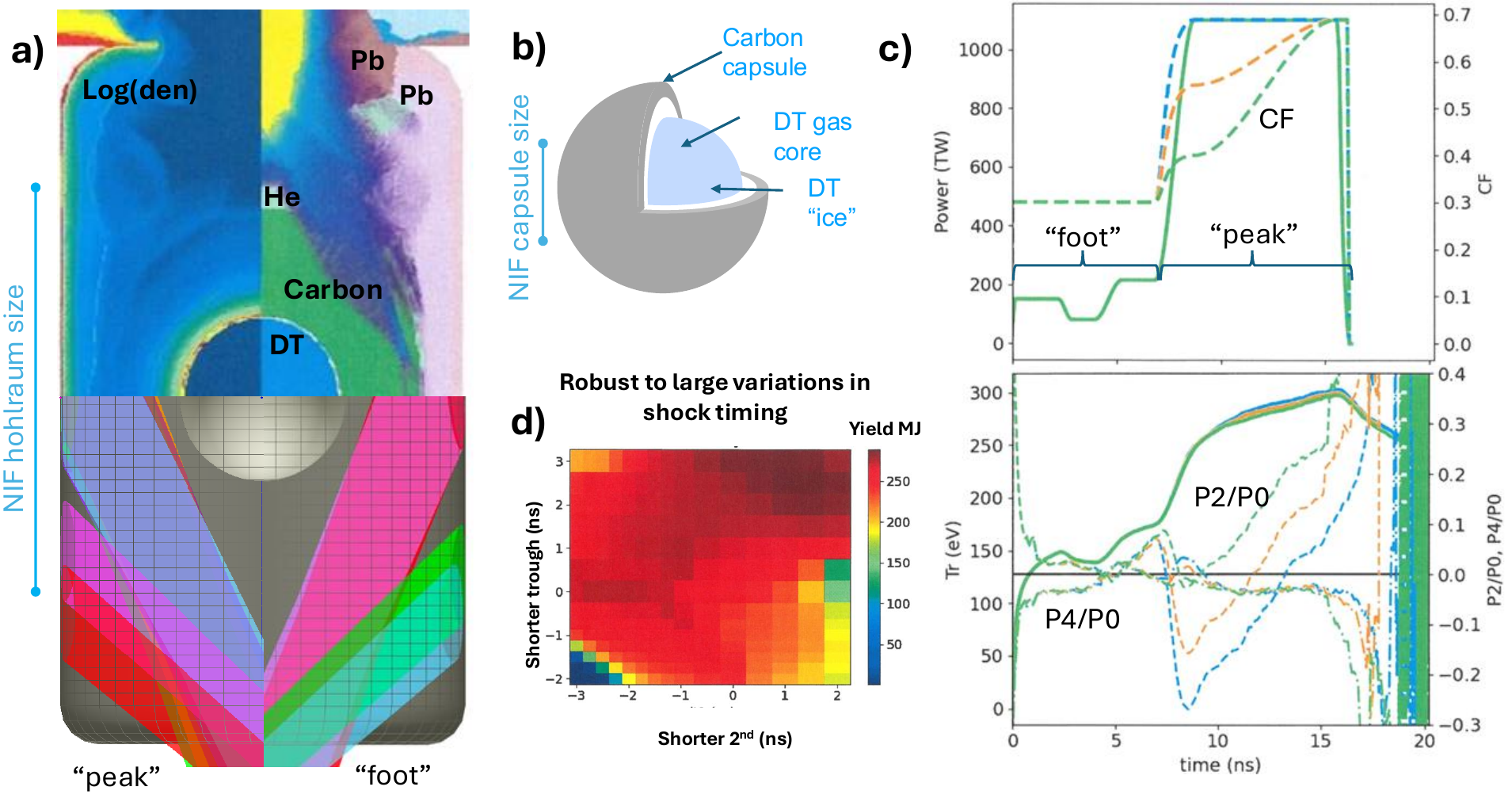}
\caption{(a) Schematic of the Inertia indirect-drive target configuration relative to the scale of current NIF ignition targets. The upper left shows the log-density distribution, the upper right shows material boundaries with laser absorption overlaid, the lower left shows the beam pointing configuration during the peak of the laser pulse, and the lower right shows the beam pointing during the foot of the pulse.  (b) Baseline capsule configuration consisting of a high-density carbon (HDC) ablator, cryogenically frozen DT fuel layer, and central DT gas region shown in scale relative to the current NIF ignition capsule size \cite{Kritcher_0808_2022,Zylstra_0808_2022}. (c) Top: example laser pulse shape and ratio of inner to total beam power (cone fraction, CF), see the text. Bottom: radiation temperature as a function of time with the corresponding P2 and P4 radiation flux asymmetries for the three cone-fraction cases. (d) Simulated target yield sensitivity to variations in shock timing, demonstrating substantial robustness to realistic timing offsets while maintaining high fusion yield.
}
\label{Fig2}
\end{figure*}

\section{LID Approach  \& Yield Scaling}
\label{Section2}
In the laser indirect-drive hot-spot ignition approach used at NIF, only a small fraction of the DT fuel is compressed and heated to fusion-relevant temperatures within a central hot spot, while a surrounding shell of lower-temperature, higher-density DT fuel (also referred to as the ``shell'') is assembled around the igniting core. If the central hot spot reaches an areal density of order $\rho R_{\mathrm{hs}}\sim0.3~\mathrm{g/cm^2}$ and a temperature of $\sim4$--$5~\mathrm{keV}$ \cite{Lindl_PoP_2004,Rosen_PoP_1999},
alpha particles produced by DT fusion reactions ($D+T \rightarrow \alpha + n$) \cite{Post_RMP_1956} become trapped within the plasma and deposit their energy locally, driving further heating of the fuel. This self-heating process leads to a rapid increase in temperature and fusion reaction rate, resulting in the thermodynamic instability known as ignition.

The implosion is initiated when high-energy laser beams are directed onto the interior of a high-$Z$ hohlraum, where the laser energy is converted into a nearly isotropic x-ray radiation field reaching temperatures of $\sim$300~eV. The x-rays ablate and compress a central capsule containing deuterium--tritium (DT) fuel in a spherically imploding geometry, where the rapid inward motion of the shell performs $PdV$ work on the central hot spot. Multiple carefully timed shocks, generated through shaping the laser power as a function of time, compress the fuel onto a low adiabat (low fuel pressure relative to the Fermi-degenerate pressure) while forming a central lower-density, higher-temperature hot spot surrounded by a dense cold fuel shell. During the implosion, the capsule radius is reduced by a factor of $\sim$25--30 and accelerated inward to velocities between 340--400~km/s, converting implosion kinetic energy into compression and heating of the central core.  These extreme velocities are required so the central hot spot ignites before energy losses in the system cool the plasma. Temperature losses occur through electron thermal conduction, bremsstrahlung x-ray emission, and hydrodynamic expansion following hot-spot formation.

When coupled to an equation for the evolution of the hot-spot mass $m$, the DT hot-spot temperature evolves according to \cite{Hurricane_PPCF_2019,Hurricane_PoP_2019}

\begin{equation}
c_{DT}\frac{dT}{dt}
=
f_{\alpha}Q_{\alpha}
-
f_{B}Q_{B}
-
Q_{e}
-
\frac{1}{m}p\frac{dV}{dt},
\label{eq:hotspot_balance}
\end{equation}
where $c_{DT}$ is the specific heat of the DT plasma, $Q_{\alpha}$ is the $\alpha$-particle heating power and $f_{\alpha}$ is the fraction deposited in the hot spot, $Q_{B}$ represents bremsstrahlung radiation losses for pure DT and $f_{B}$ is an x-ray loss factor that can increase with mix or decrease at high DT optical depth, $Q_e$ is the electron thermal conduction loss, and the final term accounts for compressional ($PdV$) work during stagnation and becomes a cooling term during expansion. A positive temperature rise ($dT/dt>0$) marks the self-heating boundary; ignition requires the trajectory to accelerate away from this boundary, often expressed as $d^2T/dt^2>0$ \cite{Ign_0808_2022}. A hot-spot areal density of $\rho R_{\mathrm{hs}}>0.3~\mathrm{g/cm^2}$ provides sufficient confinement of fusion $\alpha$ particles. At this areal density, temperatures greater than $4.3~\mathrm{keV}$ are required for $\alpha$ heating to overcome bremsstrahlung losses for pure DT; even higher temperatures are needed to offset cooling mechanisms beyond bremsstrahlung and enter a self-heating, propagating burn regime.

Following ignition, the dominant physics transitions from hot-spot formation toward propagating burn through the surrounding dense cold DT fuel layer. In this regime, alpha-particle energy deposition within the compressed fuel becomes substantially more effective, enabling more rapid propagating burn waves. To achieve high burn fractions, the thermonuclear burn wave must propagate through sufficiently dense DT fuel before hydrodynamic disassembly terminates confinement. The key scaling requirement therefore becomes maintaining sufficiently large total areal density, $\rho R_{tot}$, for efficient alpha-particle trapping and self-heating.
Thus, for larger-scale implosions with sufficient ignition margin and robustness, higher yield is achieved primarily through increased fuel burn fractions ($\sim$35--40\%) and increasing the imploded fuel mass compared to NIF rather than through higher convergence or compression.

Fusion yield ($Y$) is proportional to the product of burn fraction ($f_b$) and compressed DT fuel mass ($M_f$):

\begin{equation}
Y \sim f_b M_f,
\end{equation}
which is commonly expressed as a function of the total areal density ($\rho R_{tot}$) \cite{Fraley_PoF_1974,Christopherson_PoP_2023}:

\begin{equation}
f_b \sim \frac{\rho R_{tot}}{\rho R_{tot}+6}
\end{equation}
\\

Because the pulse is stretched in time for larger implosions, and the power required to achieve the same radiation temperature in a hohlraum scaled by $S$ scales as $S^2$, the energy scales as $E \sim S^3$. Thus, the scale factor ($S$), the fractional increase in target and capsule size, increases with laser energy ($E_d$) as:

\begin{equation}
S \sim E_d^{1/3}
\end{equation}
Then, the fuel mass $M_f$ scales like $(E_d^{1/3})^3 \sim E_d$ and the alpha-off $\rho R$ scales like $S$ so:

\begin{equation}
\label{mtv}
Y \sim {M_f} \frac{\rho R}{\rho R+6}  \sim {E_d} \frac{E_d^{1/3} }{E_d^{1/3}+6} \sim E_d^{4/3}
\end{equation}
\\

The projected fusion yield as a function of laser driver energy using the Meyer--ter--Vehn scaling law in Eq.~\ref{mtv} from NIF and NIF-EYC (Enhanced Yield Capability \cite{MacLaren_HEDP_2024}) designs is shown in Fig.~\ref{Fig1} together with current NIF experiments (blue points), optimized ``EYC'' ($E_d=2.6$~MJ) design simulations (orange points) \cite{10.1117/12.3056940}, and Inertia's design simulations (purple points).  While the idealized high-gain scaling predicts approximately $Y \propto E_d^{4/3}$, practical target designs do not scale perfectly self-similarly because important physics quantities, including opacity effects and reflected-shock strengths from the ablation front, do not scale exactly. The larger 10~MJ-class Inertia designs shown in Table~\ref{Tab1} operate at moderately higher adiabats ($\alpha \sim 3.5$--4.4) than current NIF ignition targets and employ a further $\sim$15\% increase in capsule scale beyond direct EYC scaling to increase assembled fuel mass and areal density without relying on higher convergence while remaining close to the experimentally demonstrated ignition regime. The purple points in Fig.~\ref{Fig1} correspond to Inertia designs simulated with detailed LASNEX \cite{Zimmerman1977LASNEXCF} and HYDRA \cite{Marinak_PoP_2001} radiation hydrodynamics simulations with projected yields ranging from 265--427~MJ.
Detailed integrated simulations performed on the baseline 265~MJ design achieve full yield in the presence of representative low-mode asymmetries and high-mode perturbations. The higher-yield designs maintain similar metrics important for symmetry-control (such as the length of the laser pulse) and are expected to achieve full yield with the addition of cross-beam energy transfer (CBET) and the large inner-cone power fractions available in the Inertia laser architecture. The modest increase in adiabat and higher coupled driver energy provide increased margin against hydrodynamic instability growth, shock mistiming, and target imperfections, making it particularly attractive for commercial fusion energy systems.

A key aspect of the Inertia scaling strategy is to operate sufficiently above the ignition threshold, or ``ignition cliff,'' where NIF is currently operating and where small degradations in target performance can otherwise produce disproportionately large reductions in fusion yield. Many of the non-idealities expected in a practical fusion power plant, including lower-cost materials or target fabrication processes, injection-related low-mode asymmetries, and other engineering modifications required for commercial operation, are expected to primarily shift the ignition threshold to higher required driver energies rather than lower the maximum achievable yield.  Once ignition and propagating burn are achieved, however, the larger-scale implosions with high areal densities remain capable of reaching near-full yield. By operating at higher driver energies, the Inertia designs maintain margin against these perturbations and remain on the favorable side of the ignition cliff. The integrated design calculations presented in the following section (purple points in Fig.~\ref{Fig1}) support this strategy, demonstrating high-gain operation while preserving robustness to representative perturbations and practical target design modifications.

\section{Inertia's design}
\label{Section3}
A schematic of the Inertia target and laser design is shown in Fig.~\ref{Fig2}.  The platform uses the same indirect-drive architecture as NIF, in which $\lambda_{\mathrm{laser}}\sim351$~nm laser beams are incident on the interior wall of a high-$Z$ hohlraum, generating an intense x-ray radiation bath with temperatures approaching $\sim$300~eV.  The hohlraum material is changed from gold-lined depleted uranium used at NIF to lead (Pb), a commodity material that is simpler to manufacture at scale and more compatible with high-volume fusion energy applications, while still producing a similar radiation temperature and x-ray drive spectrum to gold in ICF hohlraums \cite{Ross_SciRep_2013}.  Removing the depleted uranium slightly reduces the peak x-ray flux for a given laser power by $\sim$6.5\% \cite{Doeppner_PRL_2015}; however, the substantially higher laser power and total laser energy available in the Inertia platform compensate for this reduction.  We fill the hohlraum with a small amount of He gas (0.3~mg/cc) that has been shown to achieve effective symmetry control \cite{Kritcher_PRE_2018}  with low LPI \cite{LePape_PoP_2016, Hopkins_PRL_2015, Divol_PoP_2017} and high ($\sim$98\%) laser energy coupling to the hohlraum.  The target also includes a thin membrane, or ``tent'' \cite{Nagel_PoP_2015, Smalyuk2018HydroInstabilityReview}, used to support the capsule at the center of the hohlraum, as well as windows that maintain the hohlraum gas pressure and shield the target from heating during injection into the fusion chamber.

\begin{table}[t]
\centering
\caption{Implosion parameters for Inertia target designs compared to simulations of NIF experiment N221204~\cite{AbuShawareb2024PRL,kritcher2023} and repeats.  The ``no-$\alpha$'' yield corresponds to simulations in which alpha-particle energy deposition is disabled, preventing self-heating and propagating burn. The subscript $bn$ denotes a burn-averaged quantity. The subscript $hs$ denotes a quantity averaged over the 0.3~g/cm$^2$ hot spot.}
\label{Tab1}
\footnotesize
\setlength{\tabcolsep}{2.5pt}
\begin{tabular}{p{0.38\columnwidth}*{5}{>{\centering\arraybackslash}p{0.1\columnwidth}}}
\\
\hline
\hline
  &NIF   & Inertia ``baseline''& Inertia & Inertia & Inertia\\
\hline
Capsule I.R. (mm) 			&1.05   & 2.25 & 2.50 & 2.50 & 2.75 \\
Dopant (\%W) 				&0.6 & 0.43 & 0.50 & 0.00 & 0.50\\
Fusion yield (MJ)			& 8.0 & 265 & 353 & 296 & 427\\
No-$\alpha$ yield (MJ)		& 0.076 & 0.61 & 0.51 & 0.69 & 0.78\\
Target gain ($G$) 			&3.9& 26 & 35 & 30 & 43 \\
Laser energy ($E_d$) (MJ) 	&2.05& 10 & 10 & 10 & 10  \\
Peak laser power (PW) 		&0.44 & 1.1 & 1.1 & 1.1 & 1.1\\
Coupled energy (kJ) 			& 252 & 2476 & 2931& 2831 & 3337  \\
DT ice thickness ($\mu$m) 	&65& 126.9 & 139.8& 126.9 & 144.0 \\
DT fuel mass (mg) 			& 0.209 & 1.92 & 2.61& 2.38 & 3.27  \\
Implosion velocity (km/s) &394& 346 & 332 & 358 & 340 \\
Fuel kinetic energy (kJ)		& 16.2 & 115 & 144 & 152 & 189 \\
Adiabat 					&2.87& 3.77 & 3.73 & 3.72 & 3.66 \\
Convergence ratio 			& 22.5 & 22.2 & 19.3 & 21.4 & 19.6 \\
Coast time (ns) 			& 0.905 & 3.65 & 5.16 & 4.34 & 5.87\\
Atwood number 			&-0.061&  -0.263 &   -0.257 & -0.021 &   -0.258 \\
T$_{Ion,bn,no-\alpha }$(keV)		& 4.59 & 3.79 & 3.45& 4.22 & 3.74 \\
T$_{Ion,hs,no-\alpha }$(keV)		& 3.60 & 4.64 & 4.12 &  4.50 & 4.50\\
Total $\rho R_{no\alpha}$ ($\rho R$) (g/cm$^2$)& 1.32 & 3.31 & 3.03& 2.58 & 2.82 \\
Pressure$_{bn,no-\alpha}$ (Hot Spot, Gbar)		& 223 & 167 & 115& 116 & 107 \\
Energy$_{bn,no-\alpha}$ (Hot Spot, kJ)& 11.3 & 76.2 & 91.5 & 85.0 & 127 \\
Fuel burn fraction (\%) 		& 11.3 & 41.1 & 40.2 & 36.9 & 38.9\\
Ignition margin norm. to NIF ($EP^2$)$_{no-\alpha}$\cite{Lindl_PoP_2018,Patel_PoP_2020} & 1.0 & 3.8 & 2.2 & 2.1 & 2.6\\
 \hline
\hline
\end{tabular}
\end{table} 

\begin{figure}[t]
\includegraphics[width=.9\linewidth]{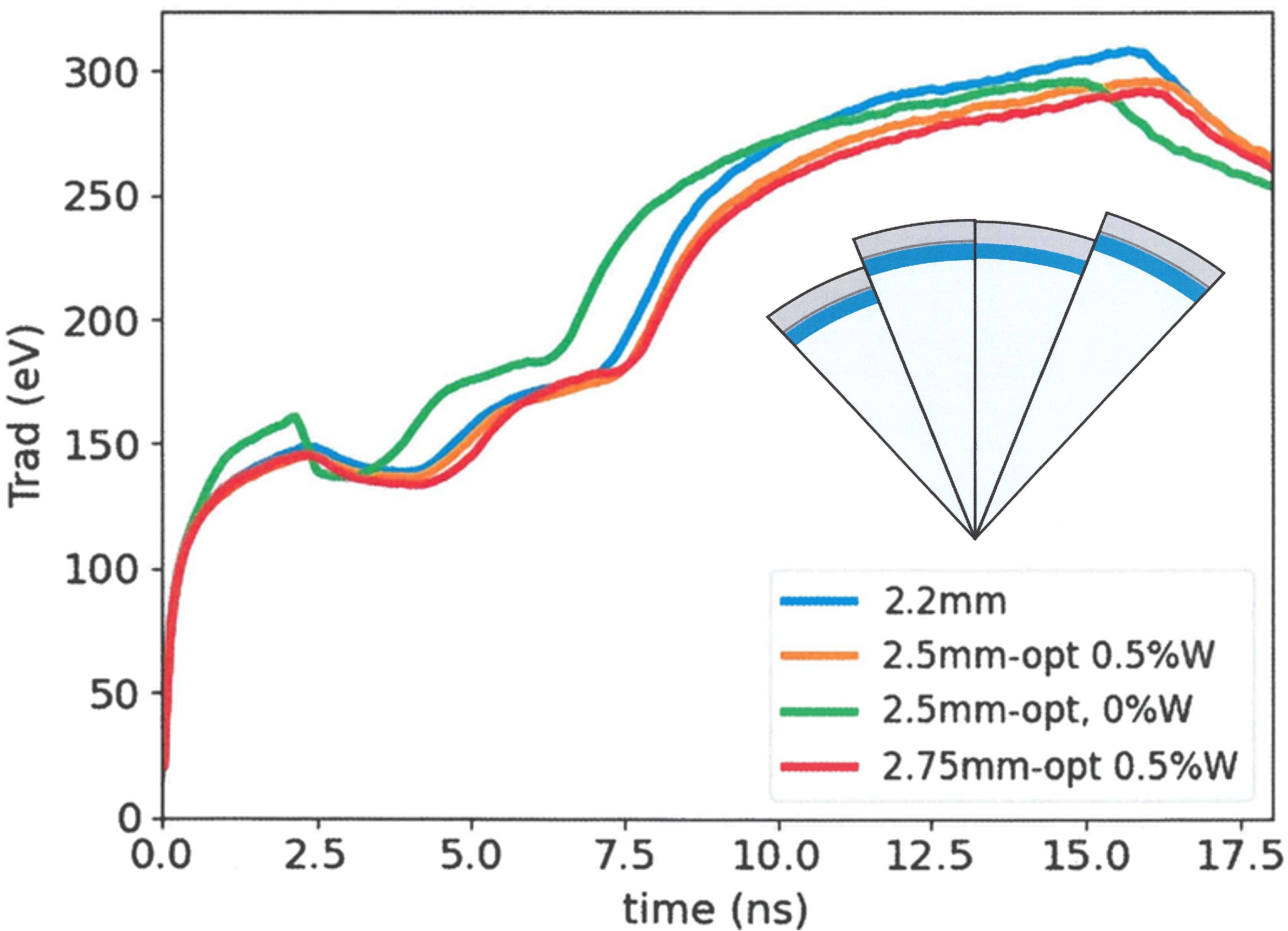}
\caption{Radiation drive as a function of time and target optimizations for the baseline and larger capsule designs, spanning 2.25--2.75~mm.  The optimization allowed modest increases in picket power, second pulse power, ice thickness, and ablator thickness while constraining the overall pulse duration to remain within $\sim$1~ns of the baseline design to maintain symmetry control.  Insets show the relative capsule scale and ablator/fuel configurations.  The optimized larger-scale capsules give substantial increases in yield, reaching $\sim$353~MJ for the 2.5~mm design and $\sim$427~MJ for the 2.75~mm design while preserving ignition-relevant implosion metrics.
}
\label{Fig3}
\end{figure}

The Inertia laser architecture employs several thousand individual laser beams, providing symmetry control and reduced sensitivity to beam-to-beam power imbalances on a given experiment. Different beam pointing configurations and power balances between beam cones (cone fractions or ``CF'') are used between the ``foot'' (0--7.5~ns) and ``peak'' (after 7.5~ns) of the pulse to maintain radiation drive symmetry control throughout the implosion, which is important for ignition and confinement, see Fig.~\ref{Fig2}(c).  A cone refers to the set of laser beams entering the hohlraum at a common polar angle relative to the hohlraum axis and the cone fraction is defined as the fraction of laser power delivered to the inner cones (21$^\circ$--33$^\circ$) aimed at the waist of the hohlraum relative to the total laser power delivered to both the inner and outer ($\sim$50$^\circ$) cones.  The nominal pointing configurations for the ``foot'' and ``peak'' of the laser pulse are shown in Fig.~\ref{Fig2}(a), bottom, the laser power and cone fractions are shown in Fig.~\ref{Fig2}(c), top, and the resulting radiation temperature and Legendre decompositions of the radiation flux asymmetry are shown in Fig.~\ref{Fig2}(c), bottom.

During the ``foot'' of the pulse, low cone fractions ($<$20\%) and beam pointings near the hohlraum P4 nodes are used to simultaneously balance early-time P4/P0 flux asymmetry and P2/P0 flux asymmetry, see Fig.~\ref{Fig2}(a), bottom right.  During the rise and ``peak'' of the laser pulse, informed by lessons learned from NIF ignition experiments, substantially higher inner-cone fractions are employed to control low-mode P2 symmetry.   See Fig.~\ref{Fig2}(a), bottom left for cone pointings during the peak of the laser pulse.  The P4 drive asymmetry smooths and its impact decreases as the capsule implodes and the outer beam spots move radially inward with hohlraum wall motion.  As the hohlraum wall ingresses into the axis of the hohlraum, a high-$Z$ ``bubble'' develops at the position of the outer beams, eventually restricting inner-beam propagation.  To compensate for this effect, the implosion is intentionally driven waist-hot ($-P_2/P_0$) during the rise and early part of the peak of the pulse, allowing the symmetry to evolve toward a pole-hot ($+P_2/P_0$) drive later in time  when the cone-fraction is no longer effective at controlling pole-waist asymmetry.  The resulting time-dependent radiation symmetry, shown in Fig.~\ref{Fig2}, is sufficient to achieve full yield in integrated LASNEX and HYDRA simulations of the baseline design (target gain $\sim 26$).  The wall motion and plasma evolution are shown in LASNEX radiation-hydrodynamics simulations in the top half of Fig.~\ref{Fig2}(a). The upper left shows the log mass density, while the upper right shows the material boundaries of the lead (Pb) hohlraum wall, capsule, and surrounding plasma with the laser deposition profile overlaid at 9.7~ns, part-way into the peak of the laser pulse.

To further optimize symmetry control and maximize laser coupling efficiency for larger-scale implosions, the design allows for the use of cross-beam energy transfer (CBET) \cite{Michel_PRL_2009,Moody_NP_2012,Kritcher_PRE_2018,Pickworth_PoP_2020}, providing an additional mechanism for symmetry tuning and risk reduction. CBET enables energy to be transferred either from the outer cones to the inner cones or vice versa, providing a powerful tuning knob to adjust the implosion symmetry toward ignition and burn conditions. This capability reduces reliance on simulation extrapolations of hohlraum wall motion for targets approximately twice the scale of current ignition experiments. To further reduce symmetry risk, the design maintains hohlraum drive and wall intensity conditions similar to those demonstrated on the NIF while distributing the energy across a substantially larger number of beams, thereby reducing the intensity carried by each individual beam.

Cross-beam energy transfer occurs when overlapping laser beams drive a plasma density perturbation due to ponderomotive pressure from the interference pattern of the overlapping beams.  Differences in the laser wavelengths of the overlapping beams change the frequency of the perturbation and the resulting refractive-index modulation produces Bragg scattering that can redirect incoming laser light from one beam to another.  By intentionally adjusting the relative laser wavelengths, the amount and direction of energy transfer can be controlled. CBET has been demonstrated extensively on the NIF and has proven to be a powerful and flexible tool for tuning implosion symmetry and redistributing cone power without significant additional laser--plasma interaction (LPI) risks.  The significantly larger beam count also reduces the intensity carried by each individual beam, helping mitigate LPI risks despite the longer plasma scale lengths and larger $f$-number geometry associated with the larger-scale hohlraum configuration.

The capsule design is chosen to remain close to the NIF ignition capsule.
It consists of a high-density carbon (HDC) ablator \cite{Dawedeit_DRM_2013,Ross_PRE_2015,LePape_PRL_2018}
surrounding a cryogenic DT ice layer and central DT vapor region, and is suspended at the center of the hohlraum by thin support membranes known as the ``tent.'' Figure~\ref{Fig2}(b) compares the capsule and fuel configuration of the current NIF ignition capsule with the Inertia design operating at more than twice the linear scale.  The HDC ablator preserves the design that enabled ignition at the NIF, also providing the strength required to survive target injection and acceleration while additionally reducing x-ray preheat of the inner HDC and cryogenic DT ice layer. 
In contrast, alternate plastic (CH) ablators that have not yet demonstrated ignition on the NIF or elsewhere would be more susceptible to heating from target injection and deformation.  The Inertia designs explore tungsten dopant concentrations inside the HDC ablator ranging from 0--0.6 at.\%, overlapping the 0.35--0.6 at.\% range fielded in NIF ignition experiments. Tungsten doping reduces preheat from high-energy ($M$-band) x-rays generated in the hohlraum, while the increased ablator thickness afforded by the larger driver energy provides additional shielding, allowing lower or zero dopant concentrations to be considered while maintaining acceptable DT fuel preheat. Lower or zero dopant concentrations may also simplify rapid HDC fabrication for IFE targets and reduce or eliminate instability growth at internal interfaces. Example ranges of HDC capsule inner radius and tungsten dopant concentration are summarized in Table~\ref{Tab1}.
 
An important feature of the Inertia design is the use of a clean cryogenic DT ice layer, preserving the dense layered fuel configuration that enabled ignition at the NIF.  To maintain a manageable on-site tritium inventory and support high-throughput target production, Inertia employs a fast-frozen polycrystalline DT ice layer that can be formed in less than three hours, rather than the high-quality single-crystal layers used in the NIF ignition experiments. While this rapid-freeze process introduces grooves, grain boundaries, and surface roughness that would significantly impact ignition-scale implosions, the larger fuel mass and increased burn margin of the Inertia design make it substantially more tolerant of these imperfections (see Section \ref{Section4}). 
Maintaining a layered cryogenic DT configuration also significantly reduces physics and engineering risk relative to alternative fuel concepts based on DT-wetted foams or premixed fuel structures. Such approaches introduce increased radiative losses \cite{Pak_PRL_2020}, higher DT vapor pressures, and potential survivability challenges during target injection, requiring larger driver energies to achieve comparable ignition conditions.  By contrast, our rapid-formation polycrystalline cryogenic DT layering approach preserves the experimentally demonstrated fuel configuration while enabling high gain and a practical path to commercial fusion energy.
 
The laser pulse shape is shown in Fig.~\ref{Fig2}(c) top, where multiple steps in power launch a sequence of shocks that compress the dense DT fuel shell on an adiabat of $3.66$--$3.77$, slightly ($\sim$1.3$\times$) higher than the ignition regime demonstrated at the NIF.  The corresponding radiation drive temperature is shown in Fig.~\ref{Fig2}(c), bottom and reaches foot and peak temperatures comparable to current NIF ignition platforms. High fusion yield is maintained across substantial variations in shock timing, demonstrating increased margin to realistic laser-pulse timing offsets, see Fig.~\ref{Fig2}(d).
The baseline design operates with an implosion velocity of $\sim$350~km/s, approximately 10\% below the 390~km/s NIF reference in Table~\ref{Tab1}, while assembling substantially larger total areal density and fuel mass at stagnation prior to the onset of strong alpha heating compared to current NIF ignition experiments, see Table~\ref{Tab1}.  Trading some implosion velocity for increased confinement and assembled fuel areal density was a key strategy enabling ignition and improved gain on the NIF and remains central to the scaling approach used here \cite{hurricane_2023,kritcher2023,Lindl_PoP_2023}.
 
In addition to the baseline design, one-dimensional optimization studies were performed to scale the $\sim$2.25~mm ignition-scale capsule to larger 2.5~mm and 2.75~mm inner radii while preserving key implosion metrics and pulse durations. Moderate adjustments to the picket power, second-pulse power, DT ice thickness, ablator thickness, and tungsten dopant concentration were permitted, while constraining the overall pulse duration to remain within $\sim$1~ns of the baseline design in order to minimize changes to symmetry control (Fig.~\ref{Fig3}). The resulting pulse shapes remain qualitatively similar across all scales, requiring only modest increases in early-time drive to maintain favorable shock timing and low-adiabat compression while preserving the symmetry-control strategy.

The optimized designs preserve the key ignition physics demonstrated at NIF while increasing ignition margin ($EP^2$, where $E$ and $P$ are the burn-averaged hot-spot internal energy and pressure) through larger fuel mass and improved confinement. The no-$\alpha$ values are evaluated with alpha-particle energy deposition disabled, so they measure the hydrodynamic assembly without burn amplification. The normalized no-$\alpha$ ignition metric ($EP^2$) remains more than a factor of two above the NIF ignition design point for all optimized designs (Table~\ref{Tab1}), indicating substantial margin prior to the onset of propagating burn. The optimized 2.5~mm and 2.75~mm designs achieve simulated fusion yields of approximately 296--427~MJ with target gains of 30--43 (Table~\ref{Tab1}), while remaining closely connected to the experimentally demonstrated ignition physics established at the NIF.

Although the increased picket power and lower convergence ratios can increase sensitivity to low-mode asymmetry, preliminary two-dimensional simulations indicate that the larger designs retain substantial robustness to low-mode perturbations. Optimization of beam pointing and power balance is ongoing to further improve symmetry control and exploit the flexibility provided by the large number of independently controlled beams. The primary new design consideration introduced by the larger capsules is the increased coast time prior to stagnation \cite{Hurricane_PoP_2017}; this interval between hohlraum cooling and peak compression could reduce the hot-spot pressure or increase instability growth. However, alternate operating points with slightly reduced drive power and longer peak pulse durations can provide a path to reducing this risk.

\begin{figure}
\centering
\includegraphics[width=.9\linewidth]{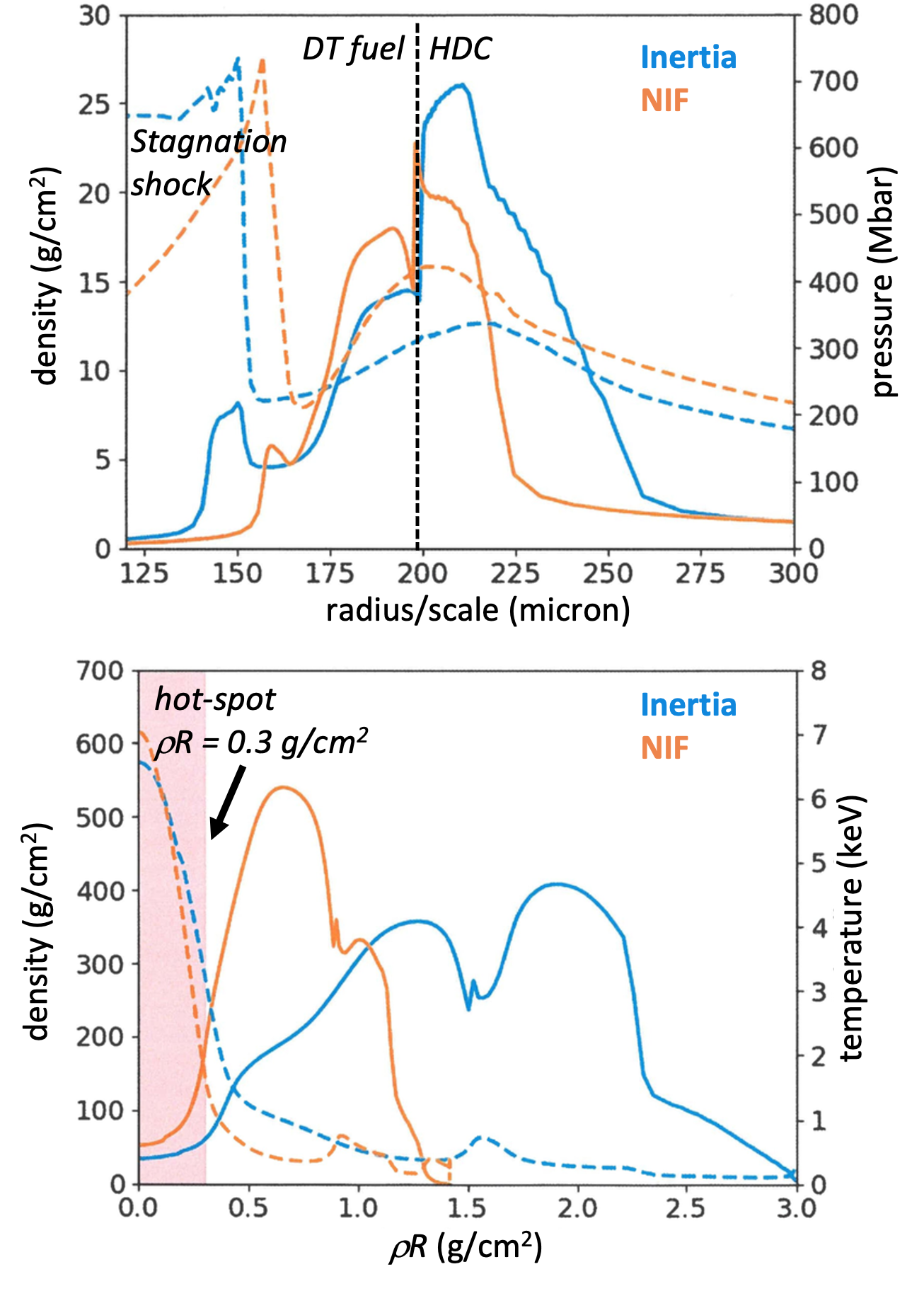}
\caption{
Radial profiles comparing the NIF ignition-scale reference capsule (orange) with the Inertia baseline design (blue). Solid curves show mass density and dashed curves show ion temperature on the corresponding axes. (a) Profiles near peak velocity across the DT-fuel/HDC-ablator interface; the Inertia design retains a thicker, cooler HDC ablator adjacent to the fuel, improving the fuel--ablator Atwood number and shielding the fuel from high-energy x-rays. (b) No-$\alpha$ bang-time profiles across the hot spot and dense fuel shell; the shaded region indicates the approximate hot-spot areal-density region for alpha-particle confinement ($\rho R_{\mathrm{hs}}\sim0.3~\mathrm{g/cm^2}$). The Inertia hot spot reaches comparable central ion temperature but a higher mass-averaged hot-spot temperature than the NIF-scale case.
}
\label{fig:density-profiles}
\end{figure}

The improved Atwood number and the effect of the longer coast time are shown in Fig.~\ref{fig:density-profiles}(a). At peak velocity, the Inertia design has substantially more ablator mass remaining. This residual mass shields the inner ablator from high-energy x-rays, keeps it cool and dense, and stabilizes the fuel--ablator Atwood number. The additional ablator mass also helps buffer the usual drawback of longer coast time: as the laser drive turns off and the hohlraum cools, the ablation pressure decreases, reducing shell density and compression at stagnation. Despite the longer coast time, the Inertia design remains approximately 25\% denser than the NIF case in the HDC region adjacent to the DT fuel.

The bang-time profiles also show that the Inertia design assembles a hotter hot spot over the alpha-stopping region than is suggested by a single averaged hot-spot temperature. Figure~\ref{fig:density-profiles}(b) shows density and ion-temperature profiles at bang time with alpha-particle deposition disabled. Hot-spot ignition requires a central region with $T_i\gtrsim4$~keV and sufficient areal density ($\rho R_{\mathrm{hs}}\sim0.3~\mathrm{g/cm^2}$) to confine and redeposit alpha-particle energy, enabling self-heating and propagating burn. Both designs reach central ion temperatures near 7~keV, but the Inertia design remains hotter across the hot-spot region, reaching a mass-averaged no-$\alpha$ hot-spot temperature of 4.64~keV compared with 3.60~keV for NIF.
A design with no tungsten dopant was also found at 2.5~mm, which may provide an easier pathway for HDC fabrication. The no-dopant design retains a stable fuel--ablator Atwood number due to the increased mass of HDC ablator in these larger-scale designs, and despite having reduced $\rho R$, it still achieves twice the no-$\alpha$ $EP^2$ of the NIF design point.

\subsection{Laser--Plasma Interactions}
Laser--plasma interactions (LPI) remain an important consideration as laser energy, hohlraum scale, and plasma path lengths increase. Critical processes include stimulated Raman scattering, stimulated Brillouin scattering, hot-electron generation, cross-beam energy transfer, and laser absorption through inverse bremsstrahlung. The baseline strategy intentionally avoids excessively aggressive laser intensity regimes and instead preserves ignition-grounded indirect-drive operating conditions while scaling to larger driver energy and hohlraum size.

Inertia’s laser architecture is designed to minimize LPI risk while enabling the substantially larger driver energies required for high-gain fusion targets. Although the total laser power increases by roughly a factor of 2--3 relative to NIF, and LPI gains scale with laser intensity ($I$) and laser wavelength ($\lambda$) as $G_{\mathrm{LPI}} \propto I \lambda^2$, Inertia's design simultaneously increases hohlraum scale, laser spot size, and total beam count. As the target scale increases by a factor $S$, the total laser power approximately scales as $S^2$, while the laser spot area also scales approximately as $S^2$, maintaining similar intensity at the hohlraum wall and LEH as current NIF ignition platforms. In addition, the number of beams increases from 192 on NIF to $\sim$1000--4000 in the Inertia architecture, reducing the intensity carried by each individual beam to roughly $1/5$--$1/20$ of NIF levels and partially offsetting the larger $f$-number geometry and longer plasma path lengths associated with larger-scale hohlraums. The platform leverages several thousand independently controlled laser beams with flexible wavelength tuning, beam pointing, SSD, bandwidth, pulse shaping, and cone fraction to maintain symmetry while reducing LPI risk.

Integrated LPI simulations performed using the LLNL parametric wave propagation code \textsc{pF3D} on the baseline high-gain hohlraum designs indicate acceptable levels of backscatter despite the increased scale of the system. Although larger hohlraums introduce longer plasma path lengths and larger effective beam $f$-number, these effects are mitigated by increasing the total number of laser beams more rapidly than the overall target energy and power scaling, thereby reducing per-beam intensity.
In parallel, Inertia plans to benchmark these models experimentally by studying beam propagation and LPI under relevant plasma conditions at smaller-scale laser facilities, while also leveraging the extensive LPI benchmarking and validation work currently being performed at the NIF.

 \section{Design robustness}
 \label{Section4}
The baseline Inertia approach preserves the ignition-proven indirect-drive approach while providing substantially greater operating margin and flexibility than ignition-threshold implosions. By operating at larger scale, higher total areal density, and higher gain, the design is significantly more tolerant of low-mode asymmetries, target imperfections, and fusion power plant constraints. This robustness enables the use of manufacturable, lower-cost targets, thicker support structures, larger holes in the capsule to leach the mandrel material and fill with fuel, commodity target materials, and lower quality polycrystalline DT ice layers, while maintaining robust ignition, propagating burn, and high thermonuclear yield. As a result, the design prioritizes robustness and commercial viability rather than maximum performance under idealized conditions.  One example of this increased margin is shown in Fig.~\ref{Fig2}, which illustrates the sensitivity of the baseline design to variations in laser shock timing. The target maintains high yield over a broad range of trough and second durations, demonstrating reduced sensitivity to laser-pulse delivery or timing errors compared to NIF experiments.

\subsection{Mode-1 Robustness}
Mode-1 asymmetry can arise from target injection errors, tracking uncertainty, beam imbalance, fuel non-uniformities, and fabrication non-uniformities. Unlike higher-order asymmetries, mode-1 produces a bulk displacement and velocity of the hot spot relative to the surrounding fuel shell, increasing residual kinetic energy (RKE) \cite{Kritcher_PoP_2014,Hurricane_PoP_2020} at stagnation and reducing confinement and alpha-particle self-heating. NIF experiments have shown that hot-spot velocity is strongly correlated with performance degradation, making mode-1 one of the most important low-mode perturbations to control.
\begin{figure}
\centering
\includegraphics[width=.9\linewidth]{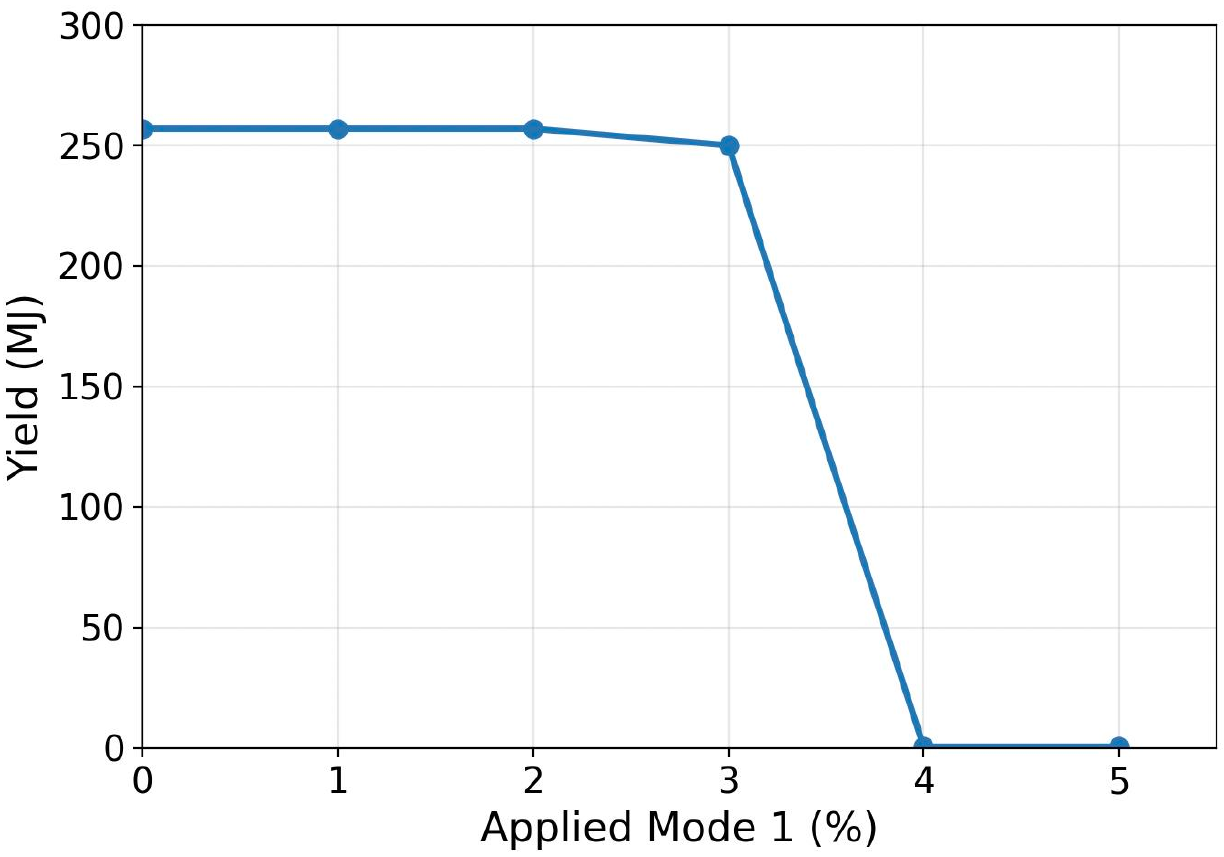}
\caption{
Fusion yield robustness for the baseline smaller-scale design under imposed mode-1 asymmetry. The target maintains high yield with applied mode-1 perturbations up to $\sim$3\%, demonstrating substantial robustness to target offset and low-mode asymmetry. Beyond this threshold, confinement rapidly degrades and the implosion fails to ignite. Relative to current NIF ignition experiments, the larger-scale Inertia platform shifts the mode-1 ignition boundary to substantially higher perturbation levels (by $\sim$3$\times$). Even greater tolerance is expected for the larger 2.5~mm and 2.75~mm capsule designs.  In practice, the combined mode-1 contributions from capsule fabrication, laser asymmetry, target injection, and tracking errors are planned to remain below this threshold.
}
\label{fig:mode1-robustness}
\end{figure}
Figure~\ref{fig:mode1-robustness} shows the sensitivity of the baseline Inertia design to imposed mode-1 perturbations. The target maintains high yield for mode-1 amplitudes up to $\sim$3\%, demonstrating substantial tolerance to target offset and low-mode asymmetry. Beyond this threshold, increasing RKE rapidly degrades confinement and ignition. Relative to NIF ignition targets, the larger fuel mass, higher gain margin, and stronger propagating burn of the Inertia platform improve mode-1 tolerance by roughly a factor of three, providing significant operating margin for fusion power plant conditions.

Estimated fusion power plant mode-1 asymmetries remain below the $\sim$3\% ignition threshold identified in Fig.~\ref{fig:mode1-robustness}. For example, a gross 20~cm target offset with laser repointing to the target produces less than 0.2\% mode-1 asymmetry, while a $\pm$200~$\mu$m laser pointing error left uncorrected produces approximately 1.5\% mode-1 using the model of MacGowan et al. \cite{MacGowan_HEDP_2020}. Target tilt is found to be negligible, with a 21~mrad tilt contributing only $\sim$0.01\% mode-1.  In the unlikely event that a large target offset occurs simultaneously with a significant laser miss-pointing error and no corrective repointing is applied, the laser may partially or completely miss the target. Such events may contribute to reduced plant operating efficiency.

\subsection{Fuel-layer quality robustness}

\begin{figure}
\centering
\includegraphics[width=1.\linewidth]{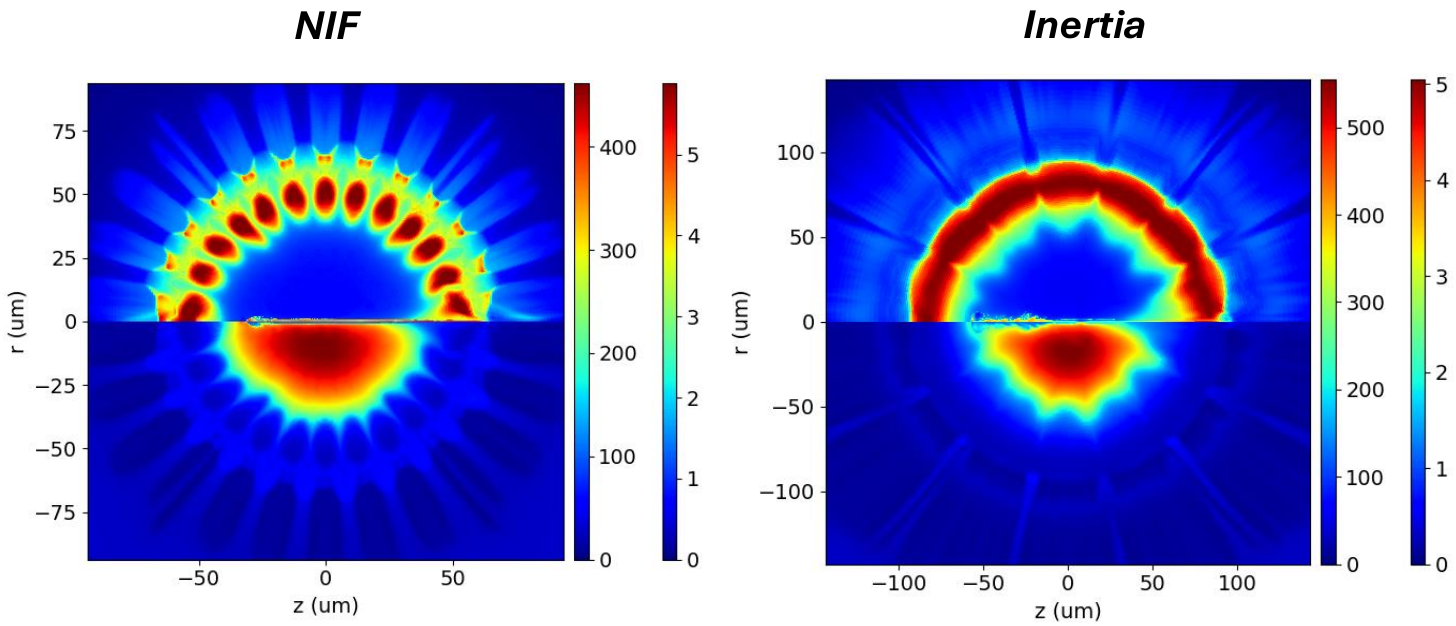}
 \caption{Simulated mass density (top, g/cm$^3$) and temperature (bottom, keV) contours at peak compression for NIF ignition-scale (left) and Inertia high-yield (right) targets, with poor-quality polycrystalline DT ice layers. The larger fuel mass and increased burn margin of the Inertia design make it substantially more resilient to dense fuel shell perturbations arising from reduced fuel-layer quality from the polycrystalline fabrication process. In this example, the Inertia target maintains robust ignition and burn propagation to full yield, while the NIF-scale target fails to ignite.}
\label{fig:ice-robustness}
\end{figure}

The larger fuel mass and higher driver energy of the Inertia design allow high yield without increasing convergence, easing Rayleigh--Taylor growth constraints and increasing tolerance to lower-quality DT ice layers, as shown in Fig.~\ref{fig:ice-robustness}. The figure compares the response of an ignition-scale NIF target and the Inertia design to perturbations arising from polycrystalline DT fuel layers.
To maintain a manageable tritium inventory and support the high target production rates required for a fusion power plant, the Inertia design uses rapidly formed polycrystalline DT ice layers that can be formed in less than three hours. Unlike the high-quality single-crystal DT layers used in NIF ignition experiments that can take days to form, this rapid-freeze process introduces grooves, grain boundaries, and surface roughness that seed hydrodynamic perturbations during the implosion.

The imposed DT-layer perturbations can significantly degrade ignition-scale implosions by reducing confinement, increasing mix, and limiting burn propagation.
Figure~\ref{fig:ice-robustness} compares an ignition-scale NIF target to the Inertia design using exaggerated perturbations from polycrystalline DT fuel layers. To amplify the effect, these simulations used 12$\times$ grooves with an initial 8~$\mu$m depth, which is approximately twice the number and depth of what is expected given experience with polycrystalline layers on the NIF. The larger Inertia capsule mitigates the impact of these initial ice grooves, retaining its full 1D yield, while the NIF implosion retains 30\% of its original performance.
The larger fuel mass, higher total areal density, and stronger alpha-heating of the Inertia design provide substantially greater tolerance to these imperfections. Although the perturbations are still visible in the stagnated shell, the design maintains robust ignition and propagating burn, allowing the target to achieve near-full yield. This increased tolerance is an important advantage for fusion energy applications, where target manufacturability, production rate, and tritium inventory are critical considerations alongside peak target performance.

\subsection{Robustness to HDC capsule defects}
Another important mass--manufacturability consideration for fusion energy targets is tolerance to defects within the high-density carbon (HDC) ablator. Small voids ($\sim5$--$80~\mu\mathrm{m}^3$) can arise during shell fabrication, as observed in capsules fielded at NIF \cite{Zylstra_PoP_2020,Zylstra_PRL_2021}, and may seed localized hydrodynamic perturbations that generate ablator jetting and mixing during the implosion. These voids arise during chemical vapor deposition (CVD) of carbon onto spherical silicon mandrels. Capsule motion within the coating chamber, required to achieve coating uniformity, can introduce isolated voids within the HDC layer \cite{Clark2024AblatorDefects,Bachmann2022DarkMix,Bachmann_PoP_2023}. Studies have shown that ablator defects can degrade performance through two distinct mechanisms: localized jets that transport ablator material into the hot spot and broader fuel--ablator mixing that reduces DT fuel compressibility and limits burn propagation. For ignition-class implosions, degradation of fuel compression can be as important as direct hot-spot contamination and should therefore be considered when establishing manufacturing requirements.

\begin{figure}[t]
\centering
\includegraphics[width=.9\linewidth]{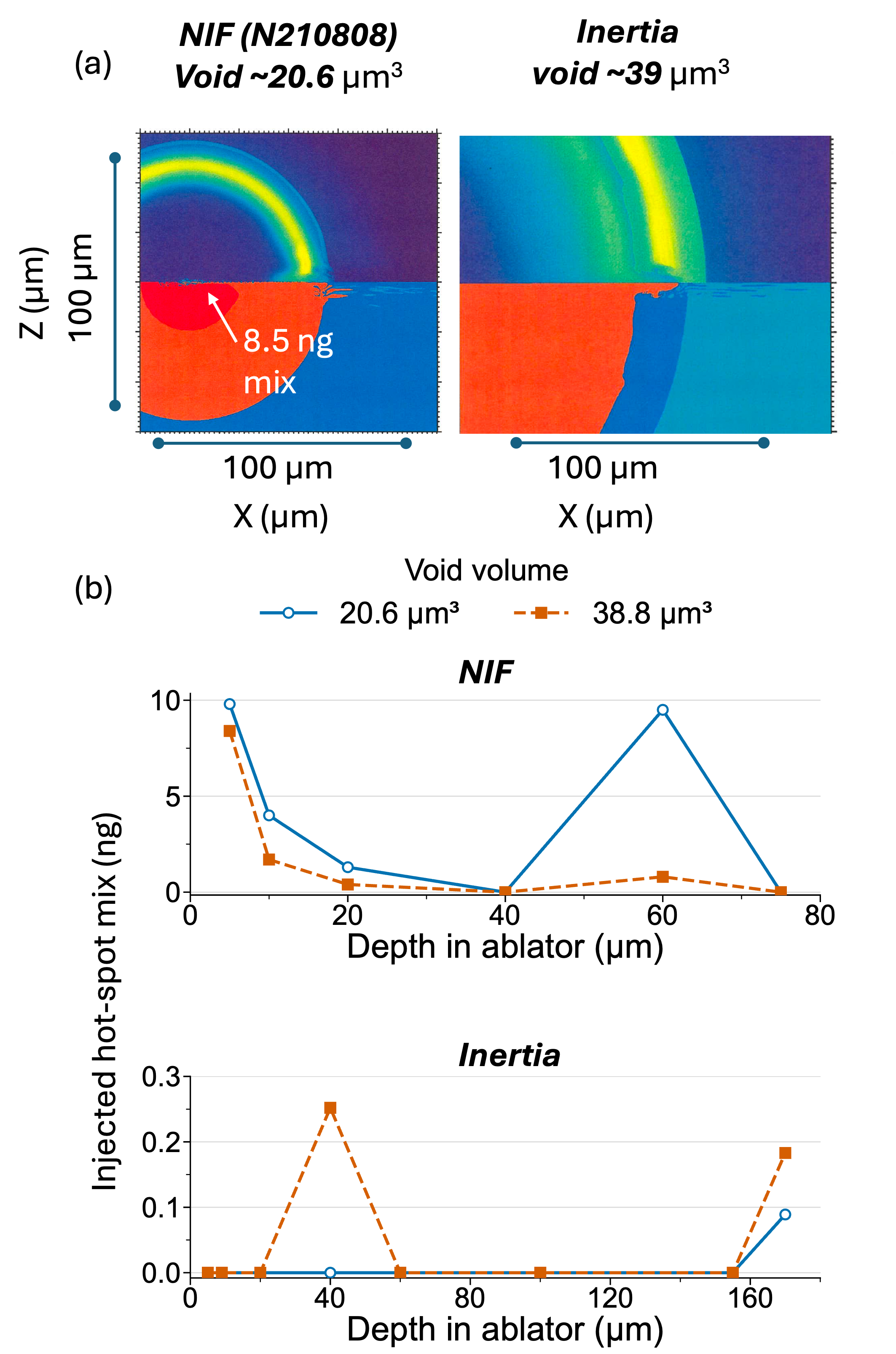}
\caption{
(a) Comparison of HDC void-induced density perturbations and ablator jetting for ignition-scale NIF capsules and the larger-scale Inertia HDC design, with density contours on the top and material boundaries on the bottom. Despite a larger void volume ($\sim$39~$\mu$m$^3$ versus $\sim$20.6~$\mu$m$^3$), the Inertia capsule does not experience mix injected into the hot spot. (b) Hot-spot mix mass as a function of HDC ablator void depth for NIF ignition-scale capsules (top) \cite{Clark2024AblatorDefects} and the larger-scale Inertia design (bottom). For comparable void volumes, the Inertia capsule generates substantially less hot-spot mix and is significantly more tolerant to void defects deeper within the ablator.
}
\label{fig:void-robustness}
\end{figure}

Figure~\ref{fig:void-robustness} compares the perturbations generated by representative HDC voids in ignition-scale NIF capsules and the larger-scale Inertia capsule design. Shown are density contours (top) and material boundaries (bottom) near peak compression for a NIF-scale capsule (left) and a 2250~$\mu$m inner-radius Inertia capsule (right). In the NIF case, the void that starts in the HDC ablator generates a jet that penetrates the hot spot and entrains 8.5 ng of ablator material into the hot DT. In the Inertia design the jet does form and no mix enters the hot spot. Only a localized density perturbation is present in the ablator. 

In the NIF calculations \cite{Clark2024AblatorDefects}, these voids were most problematic near the outside of the ablator and near the inner surface, injecting nearly 10~ng per void in the worst cases. The Inertia design also shows more sensitivity near the outside and near the doped HDC layer, but the injected mix-mass quantities are much smaller. Most void simulations resulted in no mix entering the hot spot, with three cases injecting 0.1--0.25~ng. These results demonstrate that the larger-scale Inertia platform is significantly more tolerant to manufacturing defects and void-induced perturbations than ignition-scale designs.

\begin{figure}
\includegraphics[width=.95\linewidth]{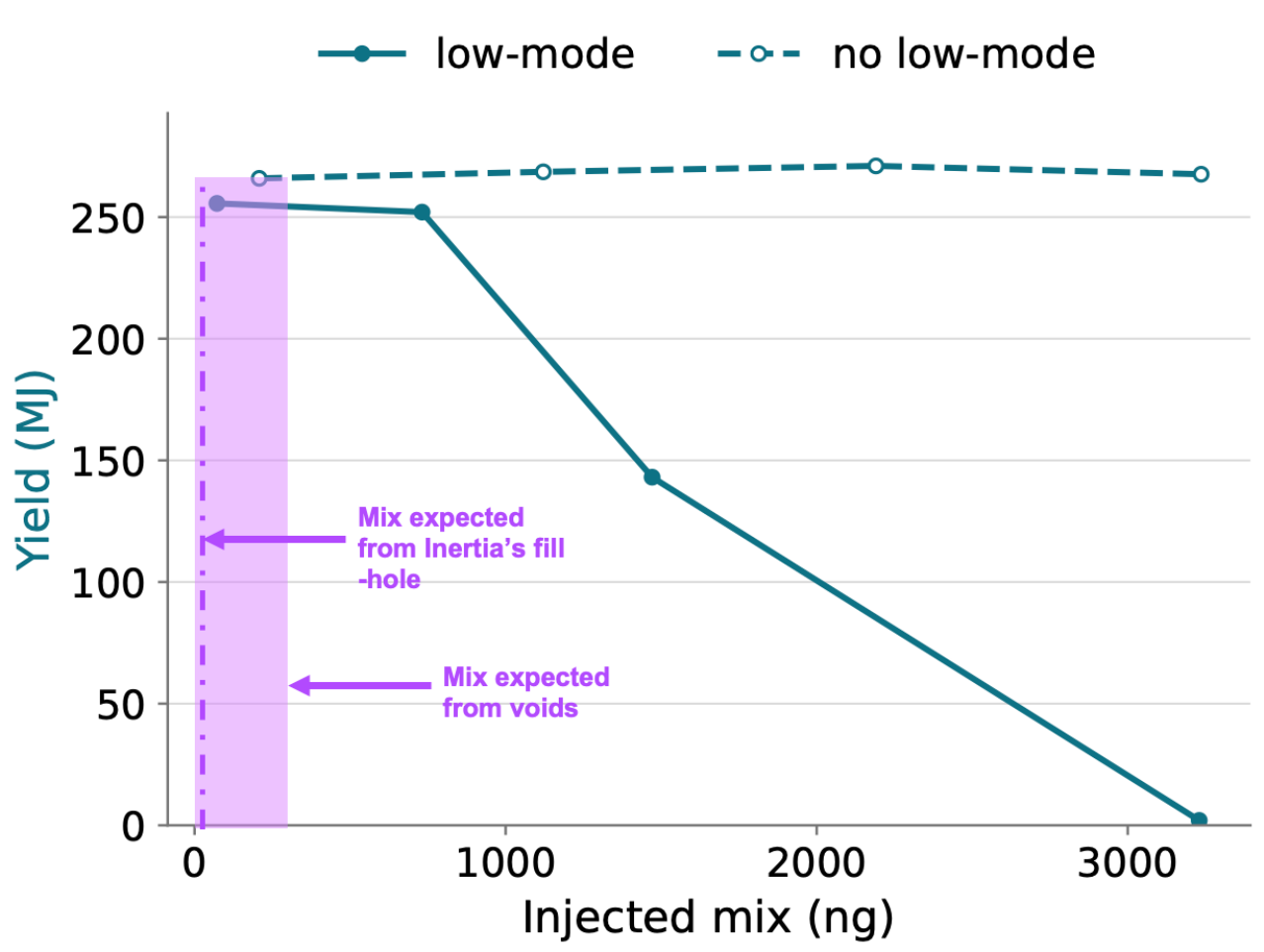}
\caption{
Performance degradation from injected hot-spot mix in the Inertia design. The implosion produces roughly the full fusion yield with up to 700 ng of injected mix when all other expected perturbations are included. When low-mode asymmetries are removed, no degradation from mix is observed through 300 ng of mix.  For Inertia's design we expect < 20 ng of mix from the fueling fill hole (dashed purple line) and expect void-induced mix to remain below our $\sim$300 ng specification for lower-cost target manufacturing (mainly voids in the shell).}
\label{fig:mix-robustness}
\end{figure}

NIF ignition experiments were found to be highly sensitive to injected mix, with 100~ng reducing yield by 30--50\% \cite{Weber2020Mixing,Divol_2023}. The Inertia design is substantially more robust, as shown in Fig.~\ref{fig:mix-robustness}. These simulations impose a surface divot of increasing size to test a range of injected-mix levels. HDC surface roughness, polycrystalline ice roughness, and the support-tent perturbation are included in both sets of simulations; the comparison differs only in whether the low-mode hohlraum-drive asymmetries are included, as discussed in the following subsection. Without the low-mode asymmetries, yield remains near 260~MJ through the largest tested mix levels. With low-mode ($P_1$--$P_6$) asymmetries from the hohlraum drive consuming some of the ignition margin, the design still tolerates up to 700~ng of injected hot-spot mix.

Several factors contribute to this improved robustness. The larger capsule scale, increased fuel and ablator thickness, lower convergence, and greater ignition margin reduce the relative impact of a fixed-size defect on the overall implosion. In addition, the higher ablator mass remaining results in a more stable fuel--ablator Atwood number than ignition-scale capsules.

These results help establish practical manufacturing specifications for fusion power plant targets. The Inertia design can tolerate lower-quality HDC capsules than are acceptable on NIF. The sensitivity studies also indicate that defects located near the outer surface of the capsule are generally more impactful than defects located deeper within the ablator. Consequently, fabrication processes can be optimized to preferentially reduce near-surface void populations, for example through longer coating times or additional processing steps.  

\subsection{Combined robustness to target modifications}

\begin{figure*}[t]
\centering
\includegraphics[width=0.8\linewidth]{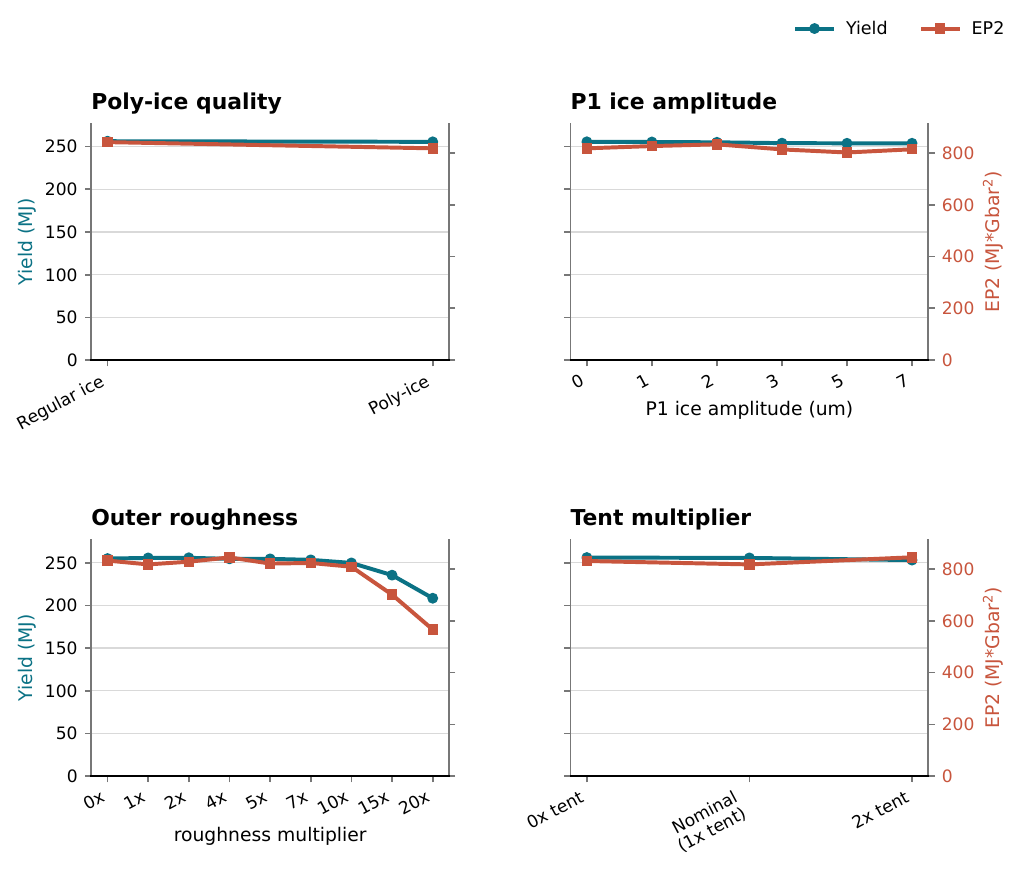}
\caption{
Integrated sensitivity of the Inertia capsule to representative power-plant target modifications. Yield (teal, left axis) and the ignition-margin metric $EP^2$ (orange, right axis) are shown for
ice model, P1 ice amplitude, outer-surface roughness multiplier,
and tent multiplier.
The target retains near-nominal yield for the nominal polycrystalline ice model, $P_1$ ice amplitudes up to 7~$\mu$m, outer roughness up to 10$\times$, and capsule-support tent perturbations up to 2$\times$ the level calculated to be required for target injection and survival. 
}
\label{fig:combined-sensitivity}
\end{figure*}

The preceding subsections considered individual perturbations such as fuel-layer roughness, capsule defects, capsule support structures capable of surviving injection into the target chamber at 10~Hz, capsule fill hole, target offsets, and low-mode asymmetries. Although these effects can be studied independently, practical fusion power plant operation will involve the simultaneous presence of multiple non-idealities. Consequently, an important requirement for a commercially relevant target platform is not simply robustness to isolated perturbations, but sufficient ignition margin to tolerate their combined effects while maintaining reliable burn propagation and high thermonuclear yield.

Figure~\ref{fig:combined-sensitivity} summarizes integrated sensitivity to target modifications around the design baseline. The baseline simulation includes low-mode radiation asymmetries arising from hohlraum dynamics, HDC surface roughness, metrologized polycrystalline ice roughness, a support-tent system designed to survive target injection, and an axial jet intended to produce fill-tube-like mix \cite{Weber2020Mixing}. 
The design appears insensitive to the polycrystalline ice model, large mode-1 ice-thickness perturbations, outer-surface roughness increases up to 10$\times$, and a support-tent perturbation 2$\times$ larger than the expected level. 
The design also remains robust to substantial hot-spot mix in the presence of the expected power-plant perturbations, maintaining high yield up to approximately 730~ng of injected ablator material. This provides substantial margin relative to the less than 20~ng of mix predicted from Inertia's fill-hole configuration and the approximately 300~ng of void-induced mix allowed by the HDC shell specification to support lower-cost manufacturing. This margin increases further as the $P_2$ and $P_4$ low-mode asymmetries are reduced.

The Inertia design philosophy intentionally prioritizes this margin. Relative to ignition-scale NIF targets, the larger fuel mass, higher assembled areal density, lower fuel--ablator Atwood number, larger burn fraction, and operation farther from the ignition threshold collectively reduce sensitivity to perturbations arising from target fabrication, injection, and laser delivery. Previous sections demonstrated substantial tolerance to low-mode asymmetries, polycrystalline DT fuel layers, and HDC void defects. Similar benefits are also simulated for other fusion power plant target modifications, including rougher capsule surfaces, thicker support structures, and capsule fill-holes optimized for manufacturability and throughput rather than minimum perturbation amplitude.

A practical fusion power plant target must survive acceleration during injection, transit through the chamber environment, and precise positioning prior to laser illumination. These requirements generally favor more robust capsule support mechanisms than those used in current ignition experiments. Likewise, rapid fueling approaches capable of filling the capsule through the HDC shell at room temperature may require larger capsule fill-holes or additional fabrication modifications compared to ignition-scale targets. Although such features introduce localized perturbations that can seed hydrodynamic growth, the larger scale and increased ignition margin of the Inertia platform substantially relax the tolerances required for successful ignition.

Taken together, power plant perturbations act primarily to reduce ignition margin rather than fundamentally alter the underlying burn physics. Once sufficient hot-spot conditions are achieved, the large assembled fuel mass and strong alpha-heating enable propagating burn to recover much of the nominal yield. This behavior is fundamentally different from ignition-threshold implosions, where relatively small degradations can move the target across the ignition cliff and produce large reductions in performance.
By operating substantially above this threshold, the Inertia platform provides the margin necessary to simultaneously accommodate a range of realistic perturbations including capsule roughness, HDC voids, polycrystalline DT ice quality, capsule support structures that withstand injection, capsule fill-hole, and low-mode asymmetries while maintaining robust ignition and high gain. This robustness is a key enabling feature for transitioning indirect-drive fusion from single-shot scientific demonstrations to practical, high-repetition-rate fusion energy systems.

\FloatBarrier
\section{Reactor-Relevant Gain and Power Production}
 \label{Section5}
\begin{table}[t]
\centering
\caption{Representative fusion power plant operating points for different target gain scenarios assuming a 10 MJ, 10 Hz driver, 12\% laser efficiency, Blanket Multiplication Factor (BMF) = 1.2, 45\% thermal-to-electric conversion efficiency, and a 50~MW$_e$ auxiliary load.}
\label{Tab2}
\small
\setlength{\tabcolsep}{3pt}
\begin{tabular}{p{3.0cm}ccc}
\\
\hline
\hline
Parameter & Gain 17 & Gain 26 & Gain 43 \\
\hline
\hline
Laser efficiency (\%) & 12 & 12 & 12 \\
Target gain ($G$) & 17 & 26 & 43 \\
Fusion yield per pulse (MJ) & 170 & 265 & 427 \\
Fusion power (MW) & 1700 & 2650 & 4270 \\
Thermal power (MW$_{th}$) & 2040 & 3180 & 5124 \\
Electric power generated (MW$_e$) & 918 & 1431 & 2306 \\
Laser recirculating power (MW$_e$) & 833 & 833 & 833 \\
Net electric to grid (MW$_e$) & 35 & 548 & 1423 \\
Process heat (MW$_{th}$) & 78 & 1218 & 3162 \\
\hline
\hline
\end{tabular}
\end{table}

The objective of high-gain target designs is to achieve reactor-relevant gains capable of supporting practical fusion power production. Higher target gain not only increases net electric output, but also reduces the recirculating power fraction and improves tolerance to fixed plant loads such as laser inefficiency, tritium processing, and auxiliary systems. As the fusion yield per shot increases, these fixed engineering overheads become a progressively smaller fraction of the total plant output, substantially improving overall plant economics and operating margin. Table~\ref{Tab2} summarizes representative operating points for an Inertia fusion power plant (FPP) employing a 10 MJ laser driver operating at 10 Hz.

Inertia's baseline target design achieves a target gain of approximately 26, producing 265~MJ of fusion energy per shot with a 10~MJ laser driver. At a repetition rate of 10~Hz, this corresponds to 2.65~GW of fusion power and 3.18~GW$_{\mathrm{th}}$ after applying a blanket multiplication factor of 1.2. Assuming a thermal-to-electric conversion efficiency of 45\% and a laser efficiency of 12\%, the plant produces approximately 1.43~GW$_e$ of gross electrical power while requiring approximately 833~MW$_e$ to operate the laser system, leaving roughly 548~MW$_e$ available for delivery to the grid.  Alternatively, instead of generating electricity, the plant can provide more than 1.2~GW$_{\mathrm{th}}$ of high-quality process heat, enabling decarbonization of industrial sectors including cement, steel, fertilizer, hydrogen production, desalination, and chemicals.  This operating point demonstrates that reactor-relevant net electric power can be achieved using a target design that remains closely connected to experimentally demonstrated NIF ignition physics rather than relying on more aggressive extrapolations.

Table~\ref{Tab2} also illustrates the impact of target gain on overall plant performance. At a lower target gain of 17, the plant remains energy positive, capable of either delivering approximately 35~MW$_e$ of net electricity or providing approximately 78~MW$_{\mathrm{th}}$ of process heat after generating sufficient electricity to satisfy the laser and auxiliary loads, and may serve as an early operating milestone before achieving the baseline target gain or further improvements in laser efficiency. At the higher-gain design point ($G\approx43$), net electric output increases to 1.42~GW$_e$ of net electricity or more than 3.1~GW$_{\mathrm{th}}$ of process heat available for industrial applications such as hydrogen production, desalination, or district heating. These examples illustrate that improvements in target gain translate directly into reduced recirculating power fraction, increased net electrical output, and greater operational flexibility for future fusion power plants.

The design philosophy presented here intentionally focuses on a baseline platform that minimizes unvalidated physics extrapolation and instead relies on controlled scaling from experimentally demonstrated designs that achieve ignition and burn, leveraging larger driver energy to simultaneously improve gain, robustness, and engineering margin for practical inertial fusion energy systems.
Future design iterations can then explore more aggressive higher-gain operating regimes that reduce laser energy requirements and incorporate advanced target concepts once the underlying physics and engineering basis has been experimentally established.

\FloatBarrier
\section{Conclusion}
 \label{Section6}
The achievement of ignition and propagating alpha-heated burn at the National Ignition Facility demonstrated that controlled thermonuclear burn is achievable in indirect-drive inertial confinement fusion and established a strong scientific foundation for inertial fusion energy. In this work, we extend this demonstrated physics to a higher-yield target platform designed to burn substantially larger fuel masses while maintaining robust implosion performance, predictive capability, and compatibility with fusion power plant requirements.

Using a 10~MJ laser driver, the designs increase DT fuel mass by approximately 9--15$\times$ relative to current NIF ignition experiments and achieve simulated fusion yields of 265--427~MJ, corresponding to target gains of 26--43. The larger fuel mass and assembled areal density of 2.6--3.3~g/cm$^2$ enable burn fractions of 37--41\% while maintaining convergence comparable to, or lower than, the NIF ignition platform. The resulting designs operate with approximately 2--4$\times$ greater modeled ignition margin than NIF, providing substantial robustness to perturbations expected in a fusion power plant environment.

These results demonstrate that target designs developed from experimentally validated NIF ignition physics can simultaneously increase fusion yield, target gain, and operating margin without requiring higher convergence or fundamentally different confinement physics. Together, these results establish a quantitative path from NIF ignition to hundreds-of-megajoule yields, target gains above 25, and the operating margin required for practical inertial fusion energy.

\section{Acknowledgements}
\label{app:collaboration}
{\bf The Inertia Collaboration:} A.~Adeloye, E.~Aguilar, M.~Albrecht, N.~Alexander, C.~Aracne, A.~Arbab, K.~Aryana, M.~Azfar, K.~Baheti, D.~Balthazor, J.~Bertsche, P.~Boca, M.~Bratsafolis, M.~Broaddus, M.~Cabral, P.~Cardamone, N.~Carlie, T.~Casserly, G.~Cearley, A.~Chilumula, T.~Clavier, J.~Dinari, A.~Dunne, S.~Ettinger, B.~Ferguson, A.~Fishkin, K.~Frick, J.~Gaffney, M.~Ganewatta, J.~Hackbarth, D.~Hammond, C.~Huang, W.~Hunder, L.~Ishigame, D.~Johnson, J.~Kenison, J.~Kim, A.~Kritcher, R.~Lau, J.~Lawson, M.~Lee, J.~Little, C.~Lim, Y.~Liu, J.~Ludwig, Z.~Mason, G.~Meric De Bellefon, S.~Pandey, J.~Peterson, A.~Piracha, P.~Planchenault, L.~Pointreau, A.~Russ, V.~Smalyuk, E.~Sprague, J.~Strickland, T.~Theiding, E.~Toro Garza, H.~Uddin, N.~Upadhye, T.~Wagner, J.~Weems, R.~Wheeler, R.~Williams, H.~Yang, and O.~Yang.  {\bf Contractors and Science and Technology Board:} M.~Fratoni, P.~Hoseman, D.~Young, L.~Suter, B.~Kruer, J.~Kilkenny, M.~Martinez, J.~Harper, A.~Bayranian, K.~Barat, P.~McKenna, J.~Zuegel, R.~Miles, S.~Burkhardt, B.~Fishler, B.~Cozad, M.~Greenberg, M.~Kepler, S.~Aracne, K.~O'Neal, J.~Wagenhoffer, S.~Aissi, M.~Adams, J.~Atherton, B.~Wirth, A.~Zude, A.~Alley, and J.~Stencil.  {\bf Lawrence Livermore National Laboratory Contractors:} S.~Chandrasekaran, J. M. Di Nicola, A.~Overland, L.~Dugan, E.~Krall, C.~Ye, M.~Triplett, N.~Hwee, F.~Fornasiero, T.~Chatterjee, S.~H.~Kim, T.~Braun, M.~Ferrucci, J.~Biener, D.~Behne, K.~Yasumura, J.~Williams, J.~McCarrick,  L.~Waxer, J.B.~McLeod, T.~Spinka, S.~Asghari, A.~Hiszpanski, N.~Ray, M.~Nelson, W.~Fenwick, and B.~Deri.

\bibliographystyle{ieeetr}
\bibliography{generated_references_cleaned}

\clearpage
\appendix
\begin{center}
{\LARGE \textbf{Appendix}}
\end{center}
\vspace{0.5cm}
\renewcommand{\thesection}{A.\arabic{section}}
\renewcommand{\thefigure}{A\arabic{figure}}
\renewcommand{\theequation}{A\arabic{equation}}

\setcounter{figure}{0}
\setcounter{equation}{0}
\section{Design Simulation Methodology}

The target designs presented in this work were developed using a combination of experimentally benchmarked multidimensional radiation-hydrodynamics simulations anchored to the National Ignition Facility (NIF) ignition results. Integrated hohlraum radiation drive, symmetry evolution, laser propagation, and implosion performance were modeled using LASNEX \cite{Zimmerman1977LASNEXCF} and HYDRA \cite{Marinak_PoP_2001}
simulations calibrated against extensive NIF experimental data. These simulations include multi-group radiation transport, laser energy deposition, non-local thermodynamic equilibrium (NLTE) atomic physics, thermal conduction, tabular equations of state, alpha-particle transport and deposition, and thermonuclear burn physics. Reduced-order view-factor integrated models were also used to rapidly explore symmetry trade spaces, cone fraction evolution, laser entrance hole motion, and hohlraum energetics, while integrated LASNEX and HYDRA simulations modeled time-dependent radiation symmetry, plasma filling, shock timing, implosion dynamics, alpha-heating, and burn propagation.  Laser--plasma interactions (LPI) were modeled using inline cross-beam energy transfer (CBET), while stimulated Raman scattering (SRS), stimulated Brillouin scattering (SBS), and laser propagation were modeled using pF3D \cite{Berger_PRL_1995,Still_PoP_2000}.

The integrated hohlraum simulations were calibrated to a suite of NIF experiments using hohlraum-wall opacity multipliers to match drive measurements of Dante hohlraum flux and capsule bang-time. This calibration set reproduces drives for 0.84 and 1.1 mm HDC capsules. Since the hohlraum conditions for this baseline design are similar to NIF, it is expected that a similar multiplier set will apply, though further experimentation will be needed. The excess $EP^2$ gives this design margin against drive losses. To assess performance with higher-resolution simulations, the frequency-dependent radiation drive from the hohlraum simulation is used to initiate higher-resolution capsule-only simulations. These drives contain the spectrally decomposed radiation asymmetries as seen by the capsule.

\section*{Disclaimer}
The opinions presented herein are those of the authors and do not necessarily reflect the opinions of the U.S. Department of Energy or the U.S. Government.

\end{document}